\documentclass[a4paper,12pt]{article}
\title {Solving the von Neumann equation with time-dependent Hamiltonian. Part I: Method}

\author{Maciej Kuna$^\dagger$ and Jan Naudts$^\circ$ \\
\strut\\
\small          $^\dagger$Wydzia\l\ Fizyki Technicznej i Matematyki Stosowanej, Politechnika Gda\'{n}ska,\\
\small	  ul. Narutowicza 11/12, 80-952 Gda\'{n}sk, Poland\\
\small	  ~~E-mail: maciek@mifgate.mif.pg.gda.pl.\\
\small          $^\circ$Departement Fysica, Universiteit Antwerpen,\\
\small          Groenenborgerlaan 171, 2020 Antwerpen, Belgium\\
\small	  ~~E-mail: jan.naudts@ua.ac.be.
}

\date {}

\usepackage{amsfonts}
\usepackage{amsmath}
\usepackage[margin=1.0 in]{geometry}

\newcommand{\be}{\begin{eqnarray}}
\newcommand{\ee}{\end{eqnarray}}

\def\Io{{\mathbb I}}
\def\Ro{{\mathbb R}}

\def\ww{v}

\newtheorem{proposition}{Proposition}{}
\newtheorem{lemma}{Lemma}{}

\def\beginproof{\par\strut\vskip 0.5cm\noindent{\bf Proof}\par}
\def\endproof{\par\strut\hfill$\square$\par\vskip 0.5cm}

\begin{document}
\maketitle

\begin{abstract}
The unitary operators $U(t)$, describing the quantum time evolution
of systems with a time-dependent Hamiltonian, can be constructed in
an explicit manner using the method of time-dependent invariants.
We clarify the role of Lie-algebraic techniques in this context
and elaborate the theory for SU(2) and SU(1,1).

We show that the constructions known as Magnus expansion and Wei-Norman expansion
correspond with different representations of the rotation group. A simpler
construction is obtained when representing rotations in terms of Euler angles.

The many applications are postponed to Part II of the paper.
\end{abstract}

\section{Introduction}

Assume that a time-dependent Hamiltonian $H(t)$ is given.
The axioms of quantum mechanics require that it is a selfadjoint operator in Hilbert space.
The solution of the time dependent Schr\"odinger equation
\be
i \frac {{\rm d}\psi}{{\rm d}t}=H(t)\psi(t)
\ee
is formally given by $\psi(t)=U(t)\psi(0)$ with unitary operators $U(t)$ satisfying
$U(0)=\Io$ and
\be
i\left(\frac {{\rm d}\,}{{\rm d}t}U(t)\right)U^*(t)=H(t).
\label {intro:ueq}
\ee
The time evolution can be extended to density operators (positive trace-class
operators with trace 1) by the relation
\be
\rho(t)=U(t)\rho(0)U^*(t).
\label {intro:timev}
\ee
They satisfy von Neumann's equations of motion
\be
\frac {{\rm d}\,}{{\rm d}t}\rho(t)=
i[\rho(t),H(t)].
\label {intro:vNeq}
\ee

Explicit expressions for $U(t)$ in function of the given $H(t)$
are usually difficult to obtain. The mathematical origin
of the difficulties is that Hamiltonians $H(t)$ and $H(t')$
at unequal times $t\not=t'$ do not necessarily
commute with each other.
Much progress has been made in the case that the time-dependent
Hamiltonian can be written as a linear combination of constant operators
$S_1, S_2, \cdots S_n$, generating a finite dimensional Lie algebra
\be
H(t)=\sum_{j=1}^nh_j(t)S_j.
\label {intro:ham}
\ee
The generators satisfy the commutation relations
\be
[S_j,S_m]=i\sum_l\omega_{jml}S_l,
\label {intro:lie}
\ee
with structure constants $\omega_{jml}$.
Then the time evolution of the generators can be written as
\be
U(t)S_jU^*(t)=\sum_mu_{mj}(t)S_m,
\label {intro:tevgen}
\ee
with coefficients $u_{mj}(t)$ solution of
\be
\frac {{\rm d}\,}{{\rm d}t}u_{mj}(t)
=\sum_{ls}\omega_{lsm}h_l(t)u_{sj}(t).
\ee
Note that (\ref {intro:tevgen}) is not the time evolution in the Heisenberg
picture. The latter is given by $U^*(t)S_jU(t)$ instead of $U(t)S_jU^*(t)$.

Using (\ref {intro:tevgen}) one can obtain the general solution of
the von Neumann equation (\ref {intro:vNeq}), with arbitrary initial conditions
of the form
\be
\rho(t=0)=C+\sum_j a_j(0)S_j.
\ee
The operator $C$ is an arbitrary operator commuting with all generators $S_j$.
The solution is
\be
\rho(t)=C+\sum_ja_j(t)S_j,
\label {intro:rhoexp}
\ee
with
\be
a_j(t)=\sum_mu_{jm}(t)a_m(0).
\ee

The problem remains to obtain an explicit expression for the
unitary operators $U(t)$.
See for instance \cite {WN63,WN64} and the references quoted there.
These authors show that one can write $U(t)$ as an ordered product
\be
U(t)=e^{g_1(t)S_1}e^{g_2(t)S_2}\cdots e^{g_n(t)S_n},
\label {intro:prodform}
\ee
where the complex functions $g_1(t),\cdots g_n(t)$ are solutions of some
set of non-linear differential equations.
In this way the problem is reduced to a well-known but rather difficult
classical (this is, non-quantum) problem of solving sets of equations.
This method has been elaborated by many authors, including \cite {LR69,GR74,TDR85,DTC86,LLMKZ96a,LLMKZ96b, SDJ97}.
A related topic is that of the superposition principle for nonlinear equations,
see for instance \cite {CGM01,CdLR08}.

Alternatively, one assumes that a particular solution $\rho_s(t)$
of von Neumann's equation (\ref {intro:vNeq}) is known.
This special solution was called a time-dependent invariant in \cite {LR69}.
We use the existence of such special solutions as a starting point
for constructing unitary operators $U(t)$, based on an idea of \cite {NK06}.


The next Section describes in general terms the method of constructing $U(t)$.
Section 3 explains how the assumption, that the Hamiltonian $H(t)$
can be expanded in terms of generators of a Lie algebra,
helps to apply the method.
Subsequent Sections specialise the method for SU(2) and SU(1,1).
The discussion of the obtained results follows in Section \ref {Sdiscus}.
In part II of the paper some applications are considered.

\section{Method}
\label {SMethod}

Assume that there is given a time-dependent Hamiltonian $H(t)$
and a special solution $\rho_s(t)$ of the von Neumann equation (\ref {intro:vNeq}).
Can one construct unitary operators $U(t)$ solving (\ref {intro:ueq})?
The method described below requires two steps to do so.

\paragraph{First step}
Determine unitary operators $V(t)$ such that
\be
\rho_s(t)=V(t)\rho_s(0)V^*(t).
\label {method:utimev}
\ee
If $\rho_s(t)$ has a discrete spectrum, as is the case for a
density operator, then this can be obtained in principle by diagonalising $\rho_s(t)$.
The unitary $V(t)$ transforms the basis in which $\rho_s(0)$ is diagonal
to the basis where $\rho_s(t)$ is diagonal.

With this $V(t)$ one can construct a Hamiltonian $K(t)$ by
\be
K(t)=i\frac {{\rm d}V}{{\rm d}t}V^*(t).
\ee
In general, this Hamiltonian differs form the given $H(t)$, in which
case $V(t)$ is not a solution of (\ref{intro:ueq}), which is the defining
equation of $U(t)$.

\paragraph{Second step}
Assume now that $\rho_s(0)$ is self-adjoint, as is the case for a density operator.
Then generically operators commuting with $\rho_s(0)$ are functions of $\rho_s(0)$
in the sense of spectral theory.
Comparing (\ref {method:utimev}) with (\ref {intro:timev}) shows that
$U(t)V^*(t)$ commutes with $\rho_s(t)$ and $V^*(t)U(t)$ commutes with $\rho_s(0)$.
Hence, one can expect that there exists a real function $f_t(x)$ such that
\be
U(t)=e^{if_t(\rho_s(t))}V(t)=V(t)e^{if_t(\rho_s(0))}.
\label {method:wshape}
\ee

It remains to determine this function $f_t(x)$.
Combine (\ref {method:wshape}) with (\ref {intro:ueq}) to obtain
\be
\frac {{\rm d}\,}{{\rm d}t}f_t(\rho_s(0))
=U^*(t)\left\{K(t)-H(t)\right\}U(t)
=V^*(t)\left\{K(t)-H(t)\right\}V(t).
\label {method:tbs}
\ee
The r.h.s.~of this equation is known from the first step of the method.
Hence the function ${\rm d}f_t(x)/{\rm d}t$ can be determined.
To this purpose, the r.h.s.~of (\ref {method:tbs}) has to be written as a
function of $\rho_s(0)$.
Next, $f_t(x)$ is obtained by integration.

\section{Lie algebras}

Assume now that $H(t)$ is a linear combination of generators $S_1,\cdots,S_n$
of a Lie algebra, with commutation relations as given by (\ref {intro:lie}), and that the
special solution $\rho_s(t)$ is of the form (\ref {intro:rhoexp}).
We first study the time evolution of the
coefficients $a_j(t)$ as elements of a Lie algebra, represented in $\Ro^n$.

Introduce the Killing form
\be
\langle a|b\rangle=\langle b|a\rangle=-\sum_{mn}\left(\sum_j\omega_{jmn}a_j\right)\left(\sum_l\omega_{lnm}b_l\right).
\ee
The minus sign is needed because of the imaginary factor $i$ in the definition (\ref {intro:lie})
of the structure constants.
Introduce the Lie bracket $a\times b=-b\times a$ by
\be
(a\times b)_l=\sum_{jm}a_jb_m\omega_{jml}.
\ee
A well-known property of the Killing form, consequence of the Jacobi identity, is that
\be
\langle a\times b|c\rangle=\langle a|b\times c\rangle.
\ee
This will be used often.

The Hamiltonian $K(t)$ can be written as
\be
K(t)=\sum_jk_j(t)S_j.
\ee
One has
\be
i\sum_j\frac {{\rm d}a_j}{{\rm d}t}S_j
&=&i\frac {{\rm d}\,}{{\rm d}t}\rho_s(t)
=[K(t),\rho_s(t)]\crcr
&=&i\sum_{jm}k_j(t)a_m(t)\omega_{jml}S_l.
\ee
This implies
\be
\frac {{\rm d}a_j}{{\rm d}t}=\sum_{ml}k_m(t)a_l(t)\omega_{mlj}.
\label {method:adot}
\ee
Introduce the notation $\dot a$ for the vector with components ${\rm d}a_j/{\rm d}t$.
From (\ref {method:adot}) then follows that $\dot a=k\times a$.

Let us assume that the Killing form is non-degenerate. This is known to be the case
if and only if the Lie algebra is semi-simple.

\begin{lemma}
Assume that the Killing form is non-degenerate.
Fix $a\in\Ro^n$ such that $\langle a|a\rangle\not=0$.
Assume that the Killing form of the Lie algebra
is such that $\langle b|a\rangle=0$ implies
that there exists $c\in\Ro^n$ such that $b=a\times c$.
Then $a\times u=0$ and $\langle u|a\rangle=0$ together imply $u=0$.
\end{lemma}

\beginproof
\nobreak

Let $d$ be arbitrary in $\Ro^n$. Write
\be
d=\frac {\langle d|a\rangle}{\langle a|a\rangle}a+b,
\ee
with $\langle b|a\rangle=0$. By assumption one can write $b=a\times c$ so that
\be
d=\frac {\langle d|a\rangle}{\langle a|a\rangle}a+a\times c.
\ee
Now calculate, using $a\times u=0$ and $\langle u|a\rangle=0$,
\be
\langle d|u\rangle = \langle a\times c|u\rangle=\langle c|u\times a\rangle = 0.
\ee
Since $d$ is arbitrary and the Killing form is non-degenerate there follows that $u=0$.

\endproof

\begin {proposition}
\label {prop1}
Let be given a special solution $\rho_s(t)$ of the form
\be
\rho_s(t)=C+\sum_j a_j(t)S_j,
\ee
where $C$ commutes with all $S_j$.
Assume that
$\langle a(0)|a(0)\rangle\not=0$
and that $\langle b|a(0)\rangle=0$ implies
that there exists $c\in\Ro^n$ such that
$b=a(0)\times c$.
Assume also that the Killing form is non-degenerate.
Then one has
\be
H(t)=K(t)+\alpha(t)\{\rho_s(t)-C\}.
\label {prop1:hk}
\ee
with
\be
\alpha(t)=\frac {\langle a|h-k\rangle}{\langle a|a\rangle}.
\label {lie:alpha}
\ee
\end {proposition}

\beginproof
Note that
\be
\dot a=h\times a=k\times a.
\ee
Let $r=h-k$. Then one has $0=r\times a$.
Let $u=r-\alpha(t)a$ with $\alpha(t)$ given by (\ref {lie:alpha}).
Then a short calculation yields $\langle u|a\rangle=0$ and $a\times u=0$.
By the previous lemma this implies that $u=0$.
Hence, (\ref {prop1:hk}) follows.

\endproof

Since $\rho_s(t)$ and $H(t)$ are known, the Proposition
ensures that the difference between $H(t)$ and $K(t)$
is proportional to the special solution $\rho_s(t)$
minus a constant operator.
Then, (\ref {method:tbs}) implies
\be
\frac {{\rm d}\,}{{\rm d}t}f_t(\rho_s(0))&=&V^*(t)\{K(t)-H(t)\}V(t)\crcr
&=&-\alpha(t)V^*(t)\{\rho_s(t)-C\}V(t)\crcr
&=&-\alpha(t)\{\rho_s(0)-C\}.
\ee
Therefore, one finds that $f_t(u)$ is the linear function given by
\be
f_t(u)=\tau(t)\,u
\quad\mbox{ with }\quad
\tau(t)=-\int_0^t{\rm d}s\,\alpha(s).
\ee
Hence, (\ref {method:wshape}) becomes
\be
U(t)=e^{i\tau(t)\rho_s(t)}V(t)=V(t)e^{i\tau(t)\rho_s(0)}.
\label {method:uvtau}
\ee
This completes the second step of the method.
All it needs from the first step is the knowledge of $V(t)$
and the value of the Killing form $\langle k|a\rangle$.

Introduce coefficients $\ww_{mj}(t)$ defined by
\be
V(t)S_jV^*(t)=\sum_m\ww_{mj}(t)S_m.
\label {method:heistev}
\ee
They satisfy
\be
a_m(t)=\sum_j\ww_{mj}(t)a_j(0).
\label {method:setofeq}
\ee
The $a_j(t)$ are known functions. We have to find $\ww_{mj}(t)$, solving (\ref {method:setofeq}),
such that unitary operators $V(t)$ exist which solve (\ref {method:heistev}).

Note that
\be
\frac {{\rm d}\,}{{\rm d}t}\langle a|a\rangle=2\langle h\times a|a\rangle=2\langle h|a\times a\rangle=0.
\ee
This means that $\langle a|a\rangle$ is constant in time.
Hence, it is obvious to look for solutions $\ww_{kj}(t)$ of (\ref {method:setofeq})
which are matrix elements of a matrix $W(t)$ that leaves the Killing form invariant,
in the sense that
\be
\langle Wb|Wc\rangle=\langle b|c\rangle
\qquad\mbox{ for all }b,c\in\Ro^n.
\label {lie:ortho}
\ee
For the construction of a matrix $W(t)$ that leaves the Killing form invariant and that satisfies (\ref {method:setofeq})
one has to rely on the large amount of knowledge about automorphisms of Lie algebras
(Note that automorphisms of the Lie algebra leave the Killing form invariant).

The next step is the construction of the operator $V(t)$.
One can make use of the fact that the construction can be done in any
faithful operator representation of the Lie algebra.
Indeed, let $\hat S_j$ be the generators in such a representation
(the hat is used to distinguish the generators $S_j$ from their representations $\hat S_j$)
and assume that
\be
\hat \rho_s(t)=\hat V(t)\hat\rho(0)\hat V^*(t)
\ee
with
\be
\hat \rho_s(t)=\sum_{j=1}^na_j(t)\hat S_j,
\ee
and
\be
\hat V(t)=e^{ir_1(t)\hat S_1}e^{ir_2(t)\hat S_2}\cdots e^{ir_n(t)\hat S_n}.
\ee
Choose for instance  a $V(t)$ of the Wei-Norman form (\ref {intro:prodform}) \cite {WN63}.
Then the operators $V(t)$, defined by
\be
V(t)=e^{ir_1(t) S_1}e^{ir_2(t) S_2}\cdots e^{ir_n(t) S_n},
\ee
satisfy
\be
\rho_s(t)= V(t)\rho(0) V^*(t).
\ee
The argument is that, when calculating $V(t)\rho(0) V^*(t)$, using the Baker-Camp\-bell-Haus\-dorff
formula, one needs to know only the commutation relations between the generators $S_j$.
Hence, the expressions for $V(t)$ and $\hat V(t)$ have the same form.
Note that, if then the generators $S_j$ are self-adjoint and the coefficients $r_j(t)$ are real,
then $V(t)$ is unitary, as requested.

Factorisations of $V(t)$, other than that of Wei and Norman, will be used below as well.
However, the above argument can be used in all cases.

\section {SU(2)}
\label {su2section}

Let be given operators $S_1,S_2,S_3$ that satisfy the commutation relations
\be
[S_1,S_2]=iS_3\quad \mbox{ and cyclic permutations}.
\label {intro:su2cr}
\ee
The non-vanishing structure constants are
$\omega_{123}=\omega_{231}=\omega_{312}=1$
and $\omega_{132}=\omega_{213}=\omega_{321}=-1$.
The Lie bracket is
\be
(a\times b)_3=a_1b_2-a_2b_1\quad \mbox{ and cyclic permutations}.
\ee
The Killing form is
\be
<a|b>=-2\sum_ja_jb_j.
\ee
A well-known identity is
\be
2x\times(x\times y)=\langle x|y\rangle x-\langle x|x\rangle y.
\label {lie:su2id}
\ee
This implies that the conditions of the Proposition \ref {prop1}
are satisfied, provided that $a(0)\not=0$. Indeed, assume $\langle b|a(0)\rangle=0$.
Then (\ref {lie:su2id}) implies that $b=a(0)\times c$ with
\be
c=-2\frac {a(0)\times b}{\langle a(0)|a(0)\rangle}.
\ee

Let be given a special solution $\rho_s(t)$ of the von Neumann equation.
Then we have to construct automorphisms $W(t)$ of the Lie algebra such that (\ref {method:utimev}) holds.
In addition, we have to calculate $\langle k|a\rangle$.
The rotations of $\Ro^3$ are the natural candidates for the $W(t)$.
Several parametrisations are possible and are treated below.

\subsection{Magnus expansion}

Any rotation of $\Ro^3$ can be denoted $R(n,\phi)$,
meaning a rotation around an axis determined by the vector $n$, satisfying $|n|=1$, and by an angle $\phi$.
One has
\be
R(n,\phi)u=\cos(\phi)u-\sin(\phi)a\times u+(1-\cos(\phi))(u\cdot n)n.
\ee
A well-known projective representation of the group $SO(3)$ in $SU(2)$ is
\be
\hat V(n,\phi)=\exp\left(-i\frac {\phi}2\sum_{\alpha=1}^3n_\alpha\sigma_\alpha\right),
\label {su2:vrep}
\ee
where the $\sigma_\alpha$ are the Pauli spin matrices. It satisfies
\be
\hat V(n,\phi)(a\cdot\sigma)\hat V(n,-\phi)
&=&\cos(\phi)a\cdot\sigma-\sin(\phi)(a\times n)\cdot\sigma\crcr
& &
+(1-\cos(\phi))(a\cdot n)n\cdot\sigma\crcr
&=&(R(n,\phi)a)\cdot \sigma.
\ee
Hence, the map $a\rightarrow \frac 12a\cdot\sigma$ is an operator representation
of the Lie algebra $SU(2)$ (note that $\sigma_j=2\hat S_j$).

Let $n(t)$ and $\phi(t)$ be smooth functions such that $a(t)=R(n,\phi)a(0)$.
Then (\ref {su2:vrep}) is an explicitly constructed representation, satisfying our requirements.
One concludes that
\be
V(t)=\exp\left(-i\phi\sum_{\alpha=1}^3n_\alpha S_\alpha\right).
\ee
This ends the first step of the method.
The disadvantage of the Magnus expansion is that the expression for the operator $K$
is rather complicated. The expansion is therefore less useful and further
application of the method is omitted here.

\subsection{Wei-Norman expansion}

A rotation in $\Ro^3$ can also be described as a sequence of rotations
about each of the principal axes. Let
\be
R_1(q_1)&=&
\left(\begin{array}{lcr}
1  &0        &0\\
0  &\cos(q_1) &\sin(q_1)\\
0  &-\sin(q_1) &\cos(q_1)
      \end{array}\right)\cr
R_2(q_2)&=&
\left(\begin{array}{lcr}
\cos(q_2)  &0 &-\sin(q_2)\\
0          &1 &0\\
\sin(q_2) &0 &\cos(q_2)
      \end{array}\right)\crcr
R_3(q_3)&=&
\left(\begin{array}{lcr}
\cos(q_3) &\sin(q_3) &0\\
-\sin(q_3) &\cos(q_3)  &0\\
0         &0          &1
      \end{array}\right).
\ee
Then one has
\be
e^{iq_1\sigma_1/2}e^{iq_2\sigma_2/2}e^{iq_3\sigma_3/2}
(u\cdot\sigma) e^{-iq_3\sigma_3/2}e^{-iq_2\sigma_2/2}e^{-iq_1\sigma_1/2}
&=&(R_1R_2R_3u)\cdot\sigma.\crcr& &
\ee
One concludes that
\be
V(t)=e^{iq_1 S_1}e^{iq_2 S_2}e^{iq_3 S_3},
\label {su2:weinorman}
\ee
with time-dependent angles $q_j(t)$ determined by
$a(t)=R_1(q_1)R_2(q_2)R_3(q_3)a(0)$.
The Hamiltonian $K$ is found to be
\be
K&=&-\dot q_1S_1-\dot q_2\{\cos(q_1)S_2-\sin(q_1)S_3\}\crcr& &
+\dot q_3\{\sin(q_2)S_1+\sin(q_1)\cos(q_2)S_2-\cos(q_1)\cos(q_2)S_3\}.\crcr& &
\ee

\subsection{Euler angles}

Another well-known parametrisation of rotations is that of Euler.
It describes the rotation of a reference frame. When
describing a rotation in $\Ro^3$
the first rotation can be used to bring the initial
vector in the plane orthogonal to direction 1.
Let
\be
a(t)=R_3(\psi)R_1(\theta)R_3(\phi)a(0).
\ee
Then one has for arbitrary $u\in\Ro^3$
\be
e^{\frac i2\psi\sigma_3}e^{\frac i2\theta\sigma_1}
e^{\frac i2\phi\sigma_3}(u\cdot\sigma)
e^{-\frac i2\phi\sigma_3}e^{-\frac i2\theta\sigma_1}
e^{-\frac i2\psi\sigma_3}
&=&(R_3(\psi)R_1(\theta)R_3(\phi)u)\cdot\sigma.\crcr
& &
\ee
One concludes that
\be
V(t)=e^{i\psi S_3}e^{i\theta S_1}e^{i\phi S_3},
\label {su2:euler}
\ee
with time-dependent Euler angles determined by
$a(t)=R_3(\psi)R_1(\theta)R_3(\phi)a(0)$.

The Hamiltonian $K$ is found to be
\be
K&=&-\dot\psi S_3-\dot\theta (\cos(\psi)S_1-\sin(\psi)S_2)\crcr
& &-\dot\phi\{\cos(\theta)S_3+\sin(\theta)\cos(\psi)S_2+\sin(\theta)\sin(\psi)S_1\}.
\ee

Note that one can replace (\ref {su2:euler}) by the alternative expressions
\be
V(t)=e^{i(\psi+\phi) S_3}e^{i\theta(\cos(\phi)S_1+\sin(\phi)S_2)}
\ee
or
\be
V(t)=e^{i\theta(\cos(\psi)S_1-\sin(\psi)S_2)}e^{i(\phi+\psi)S_3}.
\ee
These expressions might be more convenient in certain applications.

\subsection{Calculation of the time-dependent angles}

The calculation of the time-dependent angles $\psi,\theta,\phi$ may
involve complicated expressions. It is therefore noteworthy that one can
redefine the Euler angles in such a way
that (\ref {su2:euler}) may be replaced by
\be
V(t) = e^{i\phi(t)S_3}e^{i\theta(t)S_1}e^{-i\theta(0)S_1}e^{-i\phi(0)S_3}.
\label {su2:twoanglev}
\ee
This expression involves only two angles $\phi(t)$ and $\theta(t)$, determined by
\be
\sin(\phi) &=& \frac{a_1}{z}, \qquad
\cos(\phi) = \frac{a_2}{z}, \crcr
\sin(\theta) &=&-\frac{a_3}{\lambda}, \qquad
\cos(\theta) = \frac{z}{\lambda},
\ee
where $\displaystyle z(t)=\sqrt{ a_1(t)^2 + a_2(t)^2}$ and
$\displaystyle \lambda=\sqrt{z(t)^2 + a_3(t)^2}$.
To see this, note that \hfill\break
$R_1(-\theta)R_3(-\phi)$ rotates the arbitrary
vector $a$ into the fixed vector  $\lambda(0,1,0)^{\rm T}$.

Using $\displaystyle \dot\phi=\frac {a_2\dot a_1-a_1\dot a_2}{z^2}$ and $\displaystyle \dot\theta=-\frac {\dot a_3}{z}$
the Hamiltonian $K(t)$ can be calculated as follows
\be
K(t)
&=&i\frac {{\rm d}V}{{\rm d}t}V^*(t)\crcr
&=&-\dot{\theta}\{\cos(\phi)S_1 - \sin(\phi)S_2\} -
\dot{\phi}S_3 \crcr
&=& \frac{a_2\dot a_3}{z^2}S_1
-\frac{a_1\dot a_3}{z^2}S_2 + \frac{a_1\dot a_2 -a_2\dot a_1}{z^2}S_3 .
\ee
Using $\dot a=h\times a$ one sees that
this expression is of the form $k=h-\alpha(t)a$ with the function $\alpha(t)$ given by
\be
\alpha = \frac{a_1h_1 + a_2h_2}{z^2}.
\ee
The final result is then (see (\ref {method:uvtau}))
\be
U(t)=V(t)\exp\left(i\tau(t)\sum_ja_j(0)S_j\right),
\label {su2:uvtau}
\ee
with $V(t)$ given by (\ref {su2:twoanglev}) and with
\be
\tau(t)=-\int_0^t{\rm d}s\,\alpha(s)
\ee

\subsection{General solution of the equation $\dot x=h\times x$}

For the applications, discussed in Part II of this paper, it is of interest that
one can now write down the general solution of the equations $\dot x=h\times x$
in terms of the special solution $a$ and the effective time $\tau(t)$. Indeed, one has
\be
x(t)=
A\left(\begin{array}{c}
a_1\\a_2\\a_3\end{array}\right)
-\frac{\lambda B}z\cos(\lambda\tau+C)
\left(\begin{array}{c}
a_2\\-a_1\\0\end{array}\right)
+\frac Bz\sin(\lambda\tau+C)
\left(\begin{array}{c}
a_3a_1\\a_3a_2\\-z^2\end{array}\right),\crcr
& &
\label {su2:gensol}
\ee
with $A,B,C$ arbitrary constants.
The verification of this results is a matter of a straightforward calculation.

\section{SU(1,1)}
\label {Ssu11}

Consider operators $S_1,S_2,S_3$ satisfying the commutation relations
\be
[S_1,S_2]&=&iS_3\crcr
[S_2,S_3]&=&-iS_1\crcr
[S_3,S_1]&=&-iS_2.
\ee
Introduce $S'_1=iS_1$, $S'_2=iS_2$, and $S'_3=-S_3$. Then $S'_1,S'_2,S'_3$ satisfy the commutation
relations of SU(2). Hence, the results of the previous Sections can be used
to solve step 1 of our method. However, in step 2 the assumption is made that the
generators $S_1,S_2,S_3$ are self-adjoint. Therefore, the results of Section \ref {su2section}
have to be rederived here for SU(1,1).

The non-vanishing structure constants are $\omega_{123}=\omega_{321}=\omega_{132}=1$
and $\omega_{213}=\omega_{231}=\omega_{312}=-1$. The Lie bracket for vectors in $\Ro^3$ is
\be
(a\times b)_1&=&a_3b_2-a_2b_3,\crcr
(a\times b)_2&=&a_1b_3-a_3b_1,\crcr
(a\times b)_3&=&a_1b_2-a_2b_1.
\ee
The Killing form is
\be
\langle a|b\rangle=-2a_1b_1-2a_2b_2+2a_3b_3.
\label {su11:killing}
\ee
The identity (\ref {lie:su2id}) still holds. Hence, the Proposition \ref {prop1}
predicts the form of the relation between the unitary operators $V(t)$
and $U(t)$.

\subsection{Representation in $\Ro^3$}

Note that the cyclic permutation symmetry of SU(2) is lost. Hence, from the Wei-Norman
expansion (\ref {su2:weinorman}) and the expansion based on Euler angles (\ref {su2:euler})
one can derive two times three different expansions. Let us consider just one
of these, corresponding with the choice of signs (-,-,+) in the Killing form,
as found in (\ref {su11:killing}).
Then the rotations around the third axis $R_3(\phi)$ and $R_3(\psi)$ remain automorphisms. But $R_1(\theta)$
must be replaced by
\be
P_1(\chi)=\left(\begin{array}{lcr}
1   &0             &0\\
0   &\cosh(\chi)  &\sinh(\chi)\\
0   &\sinh(\chi) &\cosh(\chi)
      \end{array}\right).
\ee
Assume now that the coefficients $a_j(t)$ are written as
$a(t)=R_3(\psi)P_1(\chi)R_3(\phi)a(0)$. Then the unitary operators
$V(t)$ are given by
\be
V(t)=e^{-i\psi S_3}e^{-i\chi S_1}e^{-i\phi S_3}.
\ee
Indeed, for arbitrary $u$ in $\Ro^3$ is
\be
e^{i\psi S_3}e^{i\theta S_1}e^{i\phi S_3}
(u\cdot\sigma)
e^{-i\phi S_3}e^{-i\theta S_1}e^{-i\psi S_3}
&=&(R_3(-\psi)P_1(-\theta)R_3(-\phi)u)\cdot\sigma.\crcr
& &
\ee

The corresponding Hamiltonian is
\be
K=i\frac {{\rm d}V}{{\rm d}t}V^*(t)
&=&\dot\psi S_3+\dot\chi (\cos(\psi)S_1-\sin(\psi)S_2)\crcr
& &+\dot\phi\{\cosh(\chi)S_3+\sinh(\chi)\cos(\psi)S_2+\sinh(\chi)\sin(\psi)S_1\}.
\ee

As before in the SU(2) case one can rewrite the automorphism in
such a way that only two angles are involved
\be
a(t)=R_3(\phi(t))P_1(\chi(t)-\chi(0))R_3(-\phi(0)).
\ee
The corresponding unitary operators are
\be
V(t) = e^{-i\phi(t)S_3}e^{-i\chi(t)S_1}e^{i\chi(0)S_1}e^{i\phi(0)S_3}
\label {su11:twoanglev}
\ee
The angles $\phi(t)$ and $\chi(t)$ must satisfy
\be
\sin(\phi) &=& \frac{a_1}{z} \qquad
\cos(\phi) = \frac{a_2}{z} \crcr
\sinh(\chi) &=& \frac{a_3}{\mu} \qquad
\cosh(\chi) = \frac{z}{\mu},
\label {su11:angles}
\ee
where $\displaystyle z(t) = \sqrt{a_1(t)^2 + a_2(t)^2}$ and
$\displaystyle {\mu}= \sqrt{z(t)^2 - a_3(t)^2}$.
The fixed vector is
\be
P_1(-\chi(0))R_3(-\phi(0))a(0)=\mu(0,1,0)^{\rm T}.
\ee
Using $\dot\chi=\dot a_3/z$ and $\dot\phi=(a_2\dot a_1-a_1\dot a_2)/z^2$
the Hamiltonian K can be calculated as follows
\be
K=i\frac {{\rm d}V}{{\rm d}t}V^*(t) &=& \dot\chi\{\cos(\phi(t))S_1 - \sin(\phi(t))S_2\}
+\dot\phi S_3 \crcr
&=& \frac{\dot a_3 a_2}{z^2}S_1
-\frac{\dot a_3 a_1}{z^2}S_2
+ \frac{a_2\dot a_1-a_1\dot a_2}{z^2}S_3.
\ee
Using $\dot a=h\times a$ one sees that
this expression is of the form $k=h-\alpha a$ with the function $\alpha(t)$ given by
\be
\alpha = \frac{a_1h_1 + a_2h_2}{z^2}.
\label {su11:alphaexpr}
\ee
The final result is then (see (\ref {method:uvtau}))
\be
U(t)=V(t)\exp\left(i\tau(t)\sum_ja_j(0)S_j\right),
\label {su11:uvtau}
\ee
with $V(t)$ given by (\ref {su11:twoanglev}) and with
\be
\tau(t)=-\int_0^t{\rm d}s\,\alpha(s)
\ee

If the initial conditions satisfy $a_3^2=a_1^2+a_2^2$
then $\mu=0$ and the angle $\chi(t)$ cannot be determined by (\ref {su11:angles}).
In that case $R_3(-\phi(0))a(0)$ equals $z(0,1,1)^{\rm T}$
(we assume that $a_3(0)>0$).
The angle $\chi(0)$ can be taken equal to zero. The corresponding expression
for the time-dependent angle $\chi(t)$ is then
\be
e^{\chi(t)}=\frac {a_3(t)}{a_3(0)}.
\ee
Using $\dot\chi=\dot a_3/a_3$ and $\dot\phi=(a_2\dot a_1-a_1\dot a_2)/a_3^2$ the Hamiltonian $K$ becomes
\be
K=\frac {\dot a_3}{a_3^2}(a_2S_1-a_1S_2)+\frac {a_2\dot a_1-a_1\dot a_2}{a_3^2}S_3.
\ee
One then verifies that $k=h-\alpha a$ still holds with the function $\alpha(t)$ given by
(\ref {su11:alphaexpr}), in spite of the fact that
in this case the conditions of the Proposition \ref {prop1} are not satisfied.

Note that slightly different choices have to be made in the case
that the initial conditions satisfy $a_3^2>a_1^2+a_2^2$ instead of $a_3^2<a_1^2+a_2^2$.
Because $a_3(t)$ does not change sign, and is assumed to be positive, one can take
\be
\sinh(\chi) = \frac{z}{\mu} \qquad
\cosh(\chi) = \frac{a_3}{\mu},
\ee
with $\displaystyle {\mu}= \sqrt{a_3(t)^2-z(t)^2}$.
Then $P_1(-\chi(0))R_3(-\phi(0))a$ equals $\mu(0,0,1)^{\rm T}$.

\subsection{General solution of the equation $\dot x=h\times x$}
In the same way as for the SU(2) symmetry one can now write down the general solution
of the equation $\dot x=h\times x$. One finds, assuming $a_1^2\not=a_2^2+a_3^2$,
\be
x(t)
&=& A\left(\begin{array}{c} a_1\\a_2\\a_3\end{array}\right)
+\frac{\mu B}{z}\cosh(\mu\tau+C) \left(\begin{array}{c}
a_2\\-a_1\\0\end{array}\right) +\frac Bz\sinh(\mu\tau+C)
\left(\begin{array}{c} a_3a_1\\a_3a_2\\z^2\end{array}\right),
\label {su11:gensol}
\ee
with $A,B,C$ arbitrary constants. In the case that $a_1^2=a_2^2+a_3^2$
then one finds
\be
x(t)=(A+B\tau+C\tau^2)\left(\begin{array}{c} a_1\\a_2\\a_3\end{array}\right)
+\frac {a_3}{z^2}(B+2C\tau)\left(\begin{array}{c} a_2\\-a_1\\0\end{array}\right)
+\frac 1{z^2}C\left(\begin{array}{c} -a_1\\-a_2\\a_3\end{array}\right).
\label {su11:gensolsc}
\ee

\section{Discussion}
\label {Sdiscus}

We present a general method to construct unitary operators $U(t)$,
solving the quantum time evolution in the case of a time-dependent
Hamiltonian $H(t)$. The method starts from a special solution of
the von Neumann equation. It generalises that of \cite {LR69},
where the special solution is called a time-dependent invariant.
Here we are interested in deriving the general solution from
the knowledge of a special solution.
The method is worked out in detail for the case that $H(t)$
can be written as a linear combination of generators $S_1,\cdots,S_n$
of a Lie algebra, with time-dependent coefficients.
Most studied in the literature is the case of SU(2).
The explicit expressions for the operators $U(t)$,
known as Magnus expansion, respectively Wei-Norman expansion,
are shown to be special instances of a more general theory
and correspond to specific representations of the rotation group
in $\Ro^3$. A simpler expression for the operators $U(t)$
is obtained when the rotations are described in terms of Euler
angles -- see (\ref {su2:twoanglev}) and (\ref {su2:uvtau}).
Also the case of SU(1,1) has been discussed often.
The representation of the rotation group involving Euler angles
can be adapted to this case. The result is given by (\ref {su11:twoanglev}, \ref {su11:uvtau}).

The generator corresponding with the special solution is denoted $K(t)$.
The result of Proposition 1 is very convenient because it proves
that the difference between the Hamiltonians $H(t)$ and $K(t)$
is a linear function of the special solution $\rho_s(t)$.
The conditions of this Proposition may not always be
fulfilled. But one can fall back on the general method described in Section \ref {SMethod}.
It suffices that the spectrum of the special solution $\rho_s(t)$
is discrete and non-degenerate. Then the unitary operators
$U(t)$ and $V(t)$ are linked via (\ref {method:wshape}).
The determination of the function $f_t(x)$ is in this case only slightly more difficult
than when the Proposition can be applied.

Initially \cite {WN63,WN64}, the Wei-Norman method translated the problem
of obtaining explicit expressions for the operators $U(t)$ into sets
of differential equations with time-dependent coefficients.
In the Lie-algebraic context these equations can be written into the form
$\dot x=h\times x$, where we use the notation $h\times x$ for the Lie bracket of $h$ and $x$.
The main result of the present approach is that we obtain the general solution $x$
of these equations, starting from a special solution -- see (\ref {su2:gensol})
and (\ref {su11:gensol}, \ref {su11:gensolsc}).

The only Lie algebras discussed in detail in the present work are SU(2)
and SU(1,1). The extension to other finite-dimensional semi-simple Lie algebras
seems to be straightforward.

The Schr\"odinger and von Neumann equations with time-dependent Hamiltonian
are related to the non-linear Schr\"odinger and von Neumann equations.
In particular, a solution of the non-linear equation
\be
\frac {{\rm d}\,}{{\rm d}t}\rho(t)=i[\rho(t)^2,H]
\ee
can be used as a special solution of the von Neumann equation (\ref {intro:vNeq}) with
time-dependent Hamiltonian $H(t)=\rho(t)H+H\rho(t)$.
See for instance \cite {CDSW00}.
The theory of finding special solutions for non-linear von Neumann equations
can be found in \cite {LC98}.

\section*{}

\label{lastpage}

\end{document}